\documentclass[journal]{IEEEtran}
\usepackage{cite}
\usepackage{tabularx}
\usepackage{comment}

\ifCLASSINFOpdf
  \usepackage[pdftex]{graphicx}
\else
  \usepackage[dvips]{graphicx}
\fi
\usepackage{amsmath}
\usepackage{amssymb}
\begin{document}
%
% paper title
% Titles are generally capitalized except for words such as a, an, and, as,
% at, but, by, for, in, nor, of, on, or, the, to and up, which are usually
% not capitalized unless they are the first or last word of the title.
% Linebreaks \\ can be used within to get better formatting as desired.
% Do not put math or special symbols in the title.
\title{Dielectric Assist Accelerating Structures for Compact Linear Accelerators of Low Energy Particles in Hadrontherapy Treatments}
%
%
% author names and IEEE memberships
% note positions of commas and nonbreaking spaces ( ~ ) LaTeX will not break
% a structure at a ~ so this keeps an author's name from being broken across
% two lines.
% use \thanks{} to gain access to the first footnote area
% a separate \thanks must be used for each paragraph as LaTeX2e's \thanks
% was not built to handle multiple paragraphs
%

% Daniel de segundo y poner afiliaciones
\author{Pablo Martinez-Reviriego$^\dag$\textsuperscript{,1},~\IEEEmembership{}
        Daniel Esperante\textsuperscript{1,2},~\IEEEmembership{}
        Alexej Grudiev\textsuperscript{3},~\IEEEmembership{}
        Benito Gimeno\textsuperscript{1},~\IEEEmembership{}
        César Blanch\textsuperscript{1},~\IEEEmembership{}
        Daniel González-Iglesias\textsuperscript{1},~\IEEEmembership{}
        Nuria Fuster-Martínez\textsuperscript{1},~\IEEEmembership{}
        Pablo Martín-Luna\textsuperscript{1},~\IEEEmembership{}
        Eduardo Martínez\textsuperscript{1},~\IEEEmembership{}
        Abraham Menendez\textsuperscript{1}~\IEEEmembership{}
        and Juan Fuster\textsuperscript{1}~\IEEEmembership{}% <-this % stops a space
        \\
\textsuperscript{1}Instituto de Fisica Corpuscular (CSIC - University of Valencia), [62980] Paterna, Spain\\
\textsuperscript{2}also at Department of Electronic Engineering - ETSE, [46100] Burjassot, Spain\\
\textsuperscript{3}CERN, [1217] Meyrin, Switzerland
\thanks{$^\dag$ pablo.martinez.reviriego@ific.uv.es}}% <-this % stops a space
\maketitle

% As a general rule, do not put math, special symbols or citations
% in the abstract or keywords.
\begin{abstract}
Dielectric Assist Accelerating (DAA) structures based on ultralow-loss ceramic are being studied as an alternative to conventional disk-loaded copper cavities. This accelerating structure consists of dielectric disks with irises arranged periodically in metallic structures working under the TM$_{02}$-$\pi$ mode.
In this paper, the numerical design of an S-band DAA structure for low beta particles, such as protons or carbon ions used for Hadrontherapy treatments, is shown. Four dielectric materials with different permittivity and loss tangent are studied as well as different particle velocities.
Through optimization, a design that concentrates most of the RF power in the vacuum space near the beam axis is obtained, leading to a significant reduction of power loss on the metallic walls. This allows to fabricate cavities with an extremely high quality factor, over 100~000, and shunt impedance over 300~M$\Omega$/m at room temperature.
During the numerical study, the design optimization has been improved by adjusting some of the cell parameters in order to both increase the shunt impedance and reduce the peak electric field in certain locations of the cavity, which can lead to instabilities in its normal functioning.
\end{abstract}

% Note that keywords are not normally used for peerreview papers.

% For peer review papers, you can put extra information on the cover
% page as needed:
% \ifCLASSOPTIONpeerreview
% \begin{center} \bfseries EDICS Category: 3-BBND \end{center}
% \fi
%
% For peerreview papers, this IEEEtran command inserts a page break and
% creates the second title. It will be ignored for other modes.
\IEEEpeerreviewmaketitle

\section{Introduction}
% The very first letter is a 2 line initial drop letter followed
% by the rest of the first word in caps.
% 
% form to use if the first word consists of a single letter:
% \IEEEPARstart{A}{demo} file is ....
% 
% form to use if you need the single drop letter followed by
% normal text (unknown if ever used by the IEEE):
% \IEEEPARstart{A}{}demo file is ....
% 
% Some journals put the first two words in caps:
% \IEEEPARstart{T}{his demo} file is ....
% 
% Here we have the typical use of a "T" for an initial drop letter
% and "HIS" in caps to complete the first word.

%\cite{shintake2008compact}
\IEEEPARstart{O}{ver} many years, room-temperature disk-loaded copper radio frequency (RF) structures have been deeply studied and implemented in a wide range of applications, for fundamental science \cite{shintake2008compact} to cancer therapy \cite{tanabe1998medical} and industrial activities \cite{sethi2004electron}. However, one of the main challenges of RF cavities for accelerators consists in achieving high accelerating gradient. The Compact LInear Collider (CLIC) project \cite{aicheler2014multi} was able to reach 100 MV/m gradient for X-band normal conducting copper structures. This High Gradient (HG) technology is currently being transfered from linear colliders to different fields of application such as compact linear accelerators for Hadrontherapy treatments \cite{degiovanni2011tera, benedetti2018high, vnuchenko2020high}. Nevertheless, room temperature RF cavities are substantially less efficient than superconducting structures regarding power consumption, despite being able to reach higher gradients. An encouraging alternative to conventional disk-loaded copper structures is a dielectric loaded accelerating (DLA) structure \cite{gai1997externally, liu2003new, gold2010development}. 

A DLA structure consists of a dielectric tube surrounded by a conducting cylinder. The dielectric decreases the phase velocity  as well as the ratio of the peak electric field to the average accelerating gradient, which is about unity \cite{wei2022compact, wei2022design}. In dielectric breakdown studies, a surface field threshold of 13.8 GV/m was observed at THz frequencies \cite{thompson2008breakdown}, and no breakdowns have been observed in several high-power tests carried out on DLA structure at a level $>$5 MW. However, one of the drawbacks of dielectric structures is multipactor discharge \cite{jing2005high, jing2010progress, jing2013observation, jing2016complete}. Thus, the main issues limiting the performance of DLA structures are surface multipactor and RF breakdowns due to strong local field enhancement in micro-scale vacuum gap in the dielectric joint \cite{techaumnat2002effect, freemire2021high}.

%Consequently, the accelerating gradient achieved is potentially higher than that of disk-loaded copper structures, if metals and dielectrics had similar breakdown limit.

The concept of DLA structure was proposed in the 1940's \cite{frankel1947tm, bruck1947slow, oliner1948remarks, harvie1948proposed}, and first experimental measurements were carried out in the 1950s \cite{cohn1952design, mullett1957theoretical, walker1958vacuum}. However, disk-loaded metallic structures were more successful in that time due to their high quality factor and field holding capabilities. Recently, thanks to a remarkable progress in new ceramic materials with high dielectric permittivity ($\varepsilon_r>20$), low loss ($\tan \delta <10^{-4}$) \cite{woode1994measurement, alford1996sintered, huang2007microwave}, and ultra-low loss ($\tan\delta <10^{-5}$) \cite{templeton2000microwave, breeze2002ultralow, kanareykin2010new}, DLA structures are again being studied for multiple applications such as dual-layered DLA structure \cite{jing2008development}, a hybrid dielectric and iris-loaded accelerating structure \cite{zou2001hybrid}, and a dielectric disk accelerating (DDA) structure \cite{shao2018study}. Some examples of these dielectrics are fused silica, chemical vapor deposition (CVD) diamond or alumina, among other ceramics, some of which have been experimentally tested with high-power wakefield at Argonne National Laboratory \cite{power2004observation, jing2006observation}.

Based on these technologies, a Dielectric Assist Accelerating (DAA) structure proposed by Satoh \textit{et al.} \cite{satoh2016dielectric, satoh2017fabrication, mori2021multipactor} at C-band frequency is of particular interest since it achieved extremely high quality factor and shunt impedance. Later, this design was studied at X-band as a proposal for future linear accelerators \cite{wei2021investigations} due to its high field holding capability. Building on these developments, a DAA structure for low beta particles operating at low frequency (S-band) is studied for the first time in this work, as a solution for compact linear accelerators of low energy and low beam current, such as medical accelerators for Hadrontherapy treatments.

This paper shows a numerical study of an efficient S-band DAA structure operating under the TM$_{02}$-$\pi$ accelerating mode for four different dielectrics (CVD Diamond, MgO, MgTiO$_3$, BaTiO$_x$) and different particle velocities ($\beta~=~v/c~=~\{0.4, 0.5,...,1\}$), where $c$ is the speed of light in vacuum. This mode allows to reduce power loss on the conducting wall, achieving very high quality factor $Q_0$,
%\begin{equation}
%    Q_0=\omega\frac{W}{P_d},
%\end{equation}
and shunt impedance $Z_{eff}$,
%\begin{equation}
%    Z_{eff}=\frac{V_{acc}^2}{P_d},
%\end{equation}
at room temperature if the right dielectric material is chosen. A comparison with high-gradient copper-disk loaded structure for compact linear accelerators for medical use is shown.

% You must have at least 2 lines in the paragraph with the drop letter
% (should never be an issue)

\section{GEOMETRY OPTIMIZATION}
On the contrary to conventional disk-loaded copper structures, that operate in a TM$_{01}$ mode and  achieve high $Z_{eff}/Q_0$, DAA structures operate under a TM$_{02}$ mode in order to reduce the surface field and increase the quality factor. The dielectric allows to decrease the size of the structure and concentrate the electromagnetic energy near the beam axis, which consequently reduces copper losses and increases the shunt impedance. Thus, it is crucial that the dielectrics cost low power loss, show good thermal conductivity and withstand high electromagnetic fields. 

\begin{figure}[h]
\begin{center}
\includegraphics[width=8cm]{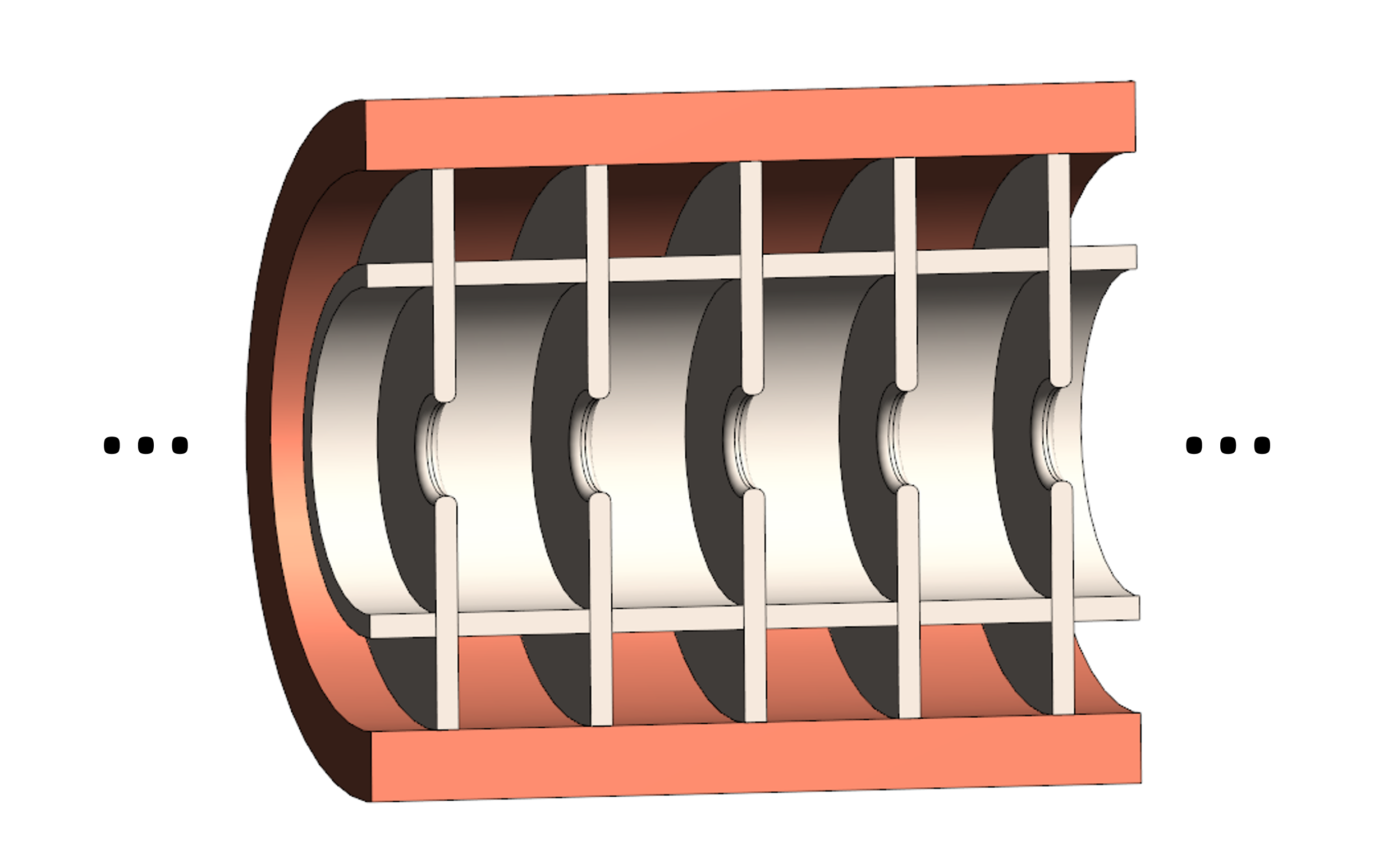}
\caption{Conceptual schematic of an infinite S-band DAA structure.}
\label{fig:structure}
\end{center}
\end{figure}

The DAA structures \cite{satoh2016dielectric, wei2021investigations} consist of axially symmetric dielectric cylinders with irises periodically arranged in a metallic enclosure operating in standing wave $\pi$-mode, as illustrated in Fig. \ref{fig:structure}. 

\subsection{Regular cell design}

\begin{figure}[h]
\begin{center}
\includegraphics[width=5cm]{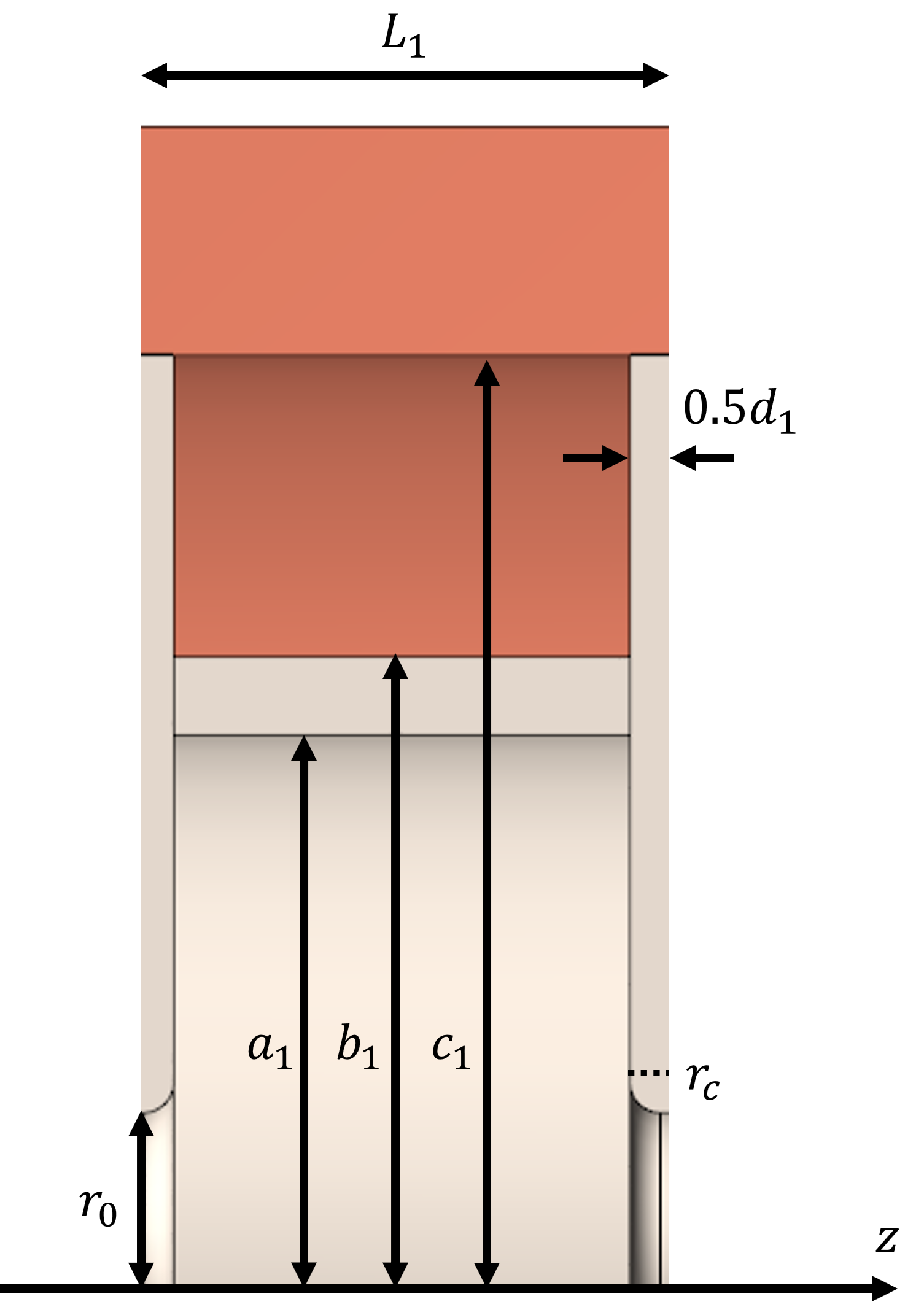}
\caption{Longitudinal cross section geometry of a regular cell of a DAA acelerator structure.}
\label{fig:cell}
\end{center}
\end{figure}

DAA design starts by optimizing parameters for the regular cell in order to maximize the $Q_0$ and the $Z_{eff}$ for the resonant frequency of interest. The longitudinal cross section of the regular cell can be seen in Fig. \ref{fig:cell}, where $r_0$ is the aperture radius, $r_c$ is the corner filet radius, $a_1$ is the inner radius, $b_1$ is the outer radius, $c_1$ is the copper waveguide radius, $d_1$ is the dielectric disk thickness, also known as iris, and $L_1$ is the constant periodic length.

Once $L_1$, $r_0$ and $r_c$ are selected based on criteria explained later, the combination of $a_1$, $b_1$, $c_1$ and $d_1$ determines the figures of merit in the cavity, such as resonant frequency, quality factor and shunt impedance of the accelerating mode TM$_{02}$-$\pi$. Thus, in this paper a new step is added to the optimization analysis, looking for the value of $d_1$ which, in combination with the three radius selection, maximizes the shunt impedance of the cavity, in spite of fixing this value to $d_1=\lambda_0/(4\sqrt{\varepsilon_r})$ as was done in previous studies \cite{satoh2016dielectric, wei2021investigations}. Thanks to the axial symmetry, optimum parameters can be calculated using the SUPERFISH tool \cite{superfish}, in addition results have been cross checked using HFSS \cite{HFSS}. Periodic boundary conditions were applied to regular cell in order to simulate an infinite long structure.

\begin{figure}[h!]
\begin{center}
\includegraphics[width=0.5\textwidth]{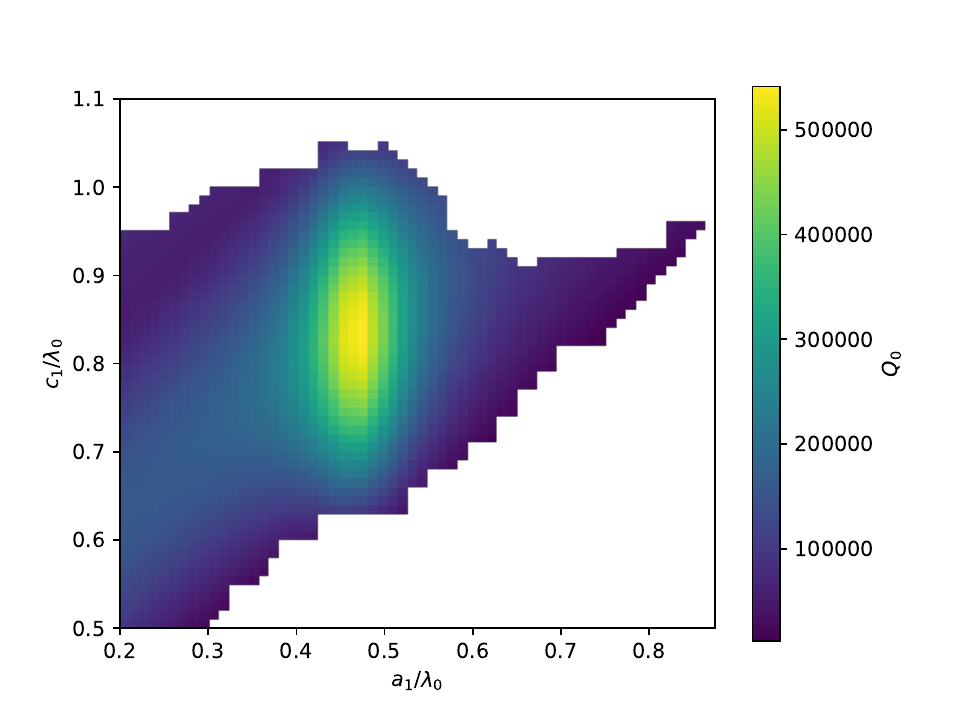}
\includegraphics[width=0.5\textwidth]{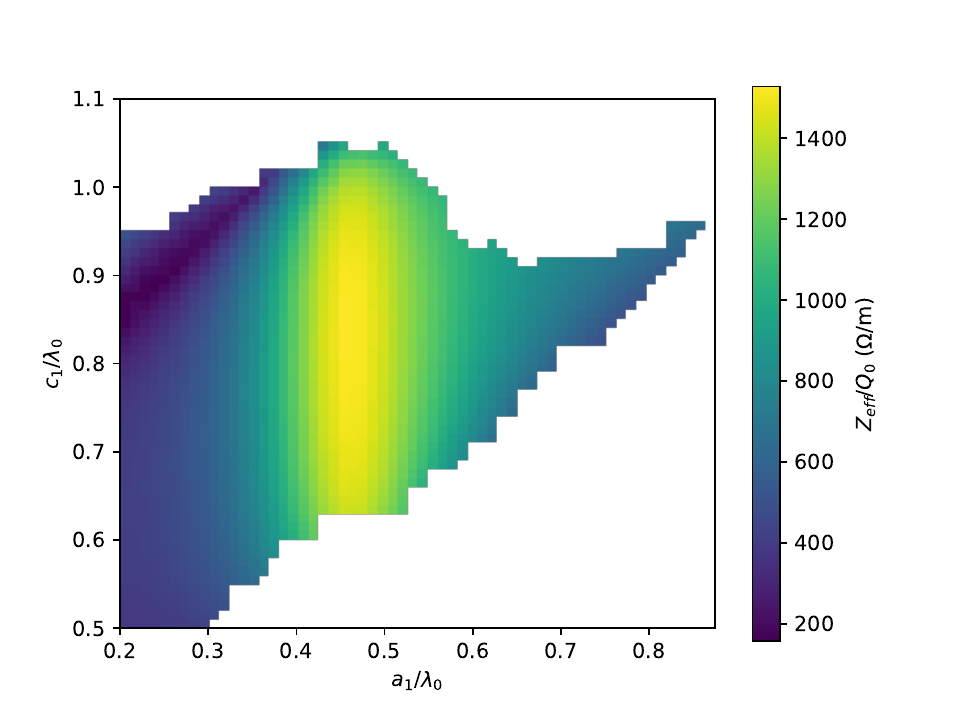}
\includegraphics[width=0.5\textwidth]{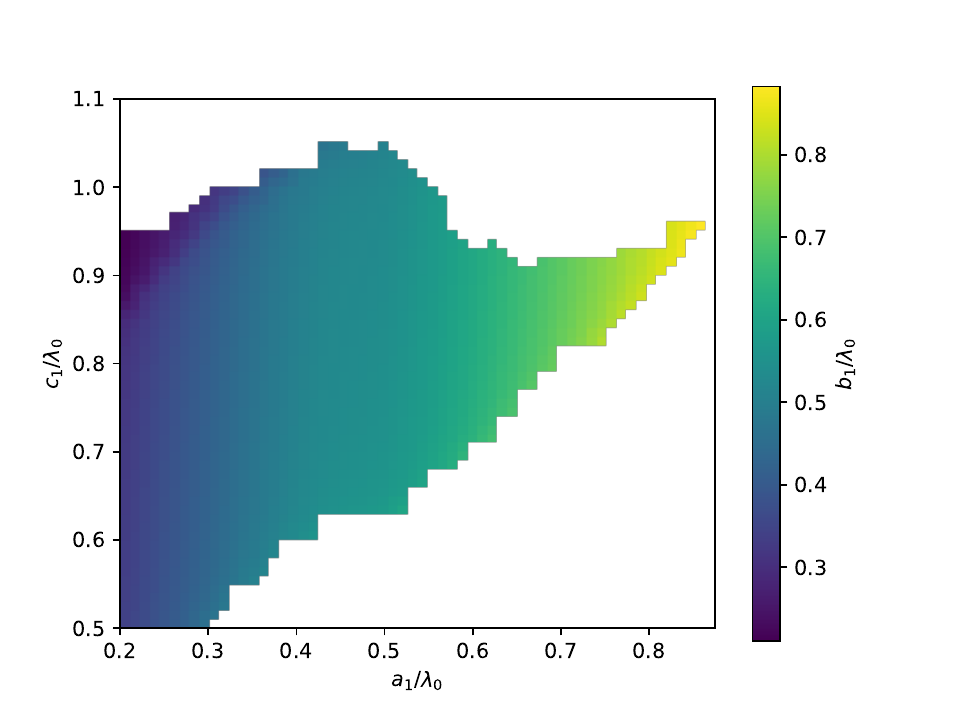}
\caption{Numerical unloaded quality factor $Q_0$, shunt impedance over quality factor $Z_{eff}/Q_0$ and geometric parameter $b_1$ as a function of geometrical parameters $a_1$ and $c_1$ for a regular cell using ideal dielectric MgTiO$_3$ and $\beta~=~0.6$. White region is due to the absence of a valid solution for the geometry.}
\label{fig:optimization_1}
\end{center}
\end{figure}

The resonant frequency goal was fixed at $f_0=(3000\pm 2)$ MHz, $L_1=\beta \lambda_0/2$, where $\lambda_0=c/f_0$ is the free space wavelength, $r_0=2$ mm for comparison with high gradient copper structures and $r_c=d_1/2$. Then, in order to find the best values for $d_1$ and the radii $a_1, b_1, c_1$, a two step scan needs to be done. First, $d_1$ is fixed at its initial value $d_1=d_0=\lambda_0/(4\sqrt{\varepsilon_r})$, so the resonant frequency $f_0$ is determined by the combination of $a_1$, $b_1$ and $c_1$. Once $a_1$ and $c_1$ are fixed, the value of $b_1$ can be recalculated for the given frequency $f_0$ using the numerical solver. Following this methodology, optimum parameters can be found by sweeping through $a_1$ and $c_1$, finding the corresponding value of $b_1$, $Q_0$ and $Z_{eff}/Q_0$. This is shown in Fig. \ref{fig:optimization_1} for MgTiO$_3$ with $\varepsilon_r=16.66$ and $\tan\delta=0$. It must be noted here that for low electric permittivity and low particle velocity, it might be impossible to find a geometry of a regular cell with the desired resonant frequency.

\begin{figure}[h]
\begin{center}
\includegraphics[width=0.5\textwidth]{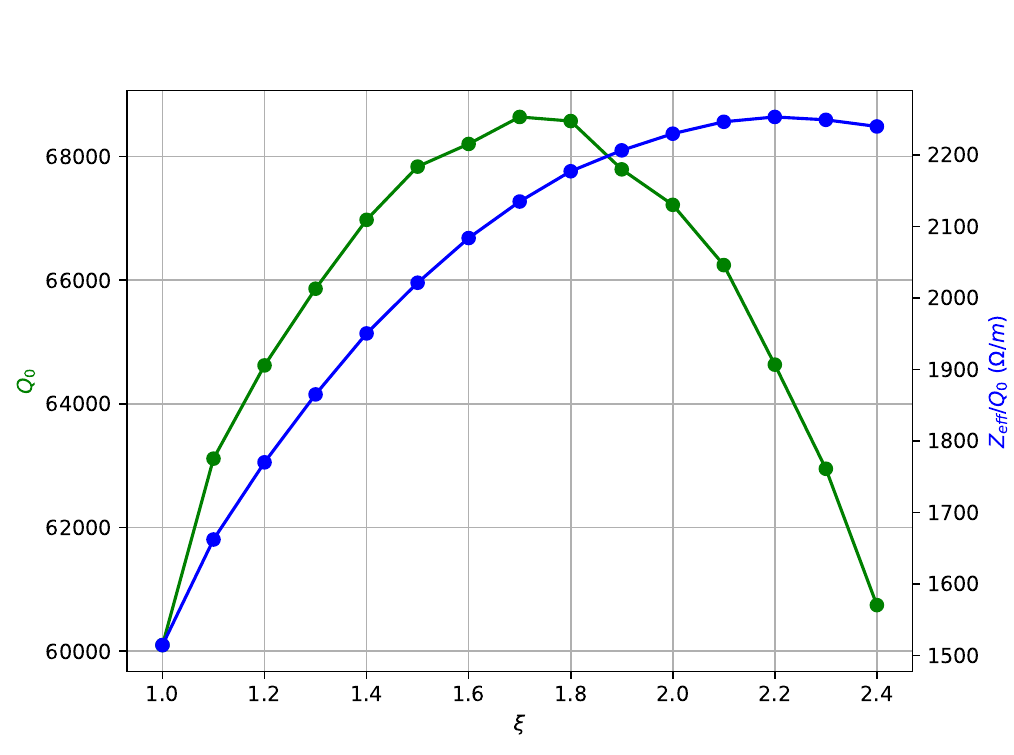}
\caption{Numerical unloaded quality factor $Q_0$ and shunt impedance over quality factor $Z_{eff}/Q_0$ as a function of normalized iris thickness for a regular cell using real dielectric MgTiO$_3$ and $\beta=0.6$.}
\label{fig:optimization_2}
\end{center}
\end{figure}

This process is repeated for each value of a second swept in $d_1=\xi d_0$, where $\xi$ is the normalized iris thickness. This allows to find a better solution in terms of $Z_{eff}$, as illustrated in Fig. \ref{fig:optimization_2} for MgTiO$_3$ with $\varepsilon_r=16.66$ and $\tan\delta=3.43\times10^{-5}$.

\begin{figure}[h]
\begin{center}
\includegraphics[width=8cm]{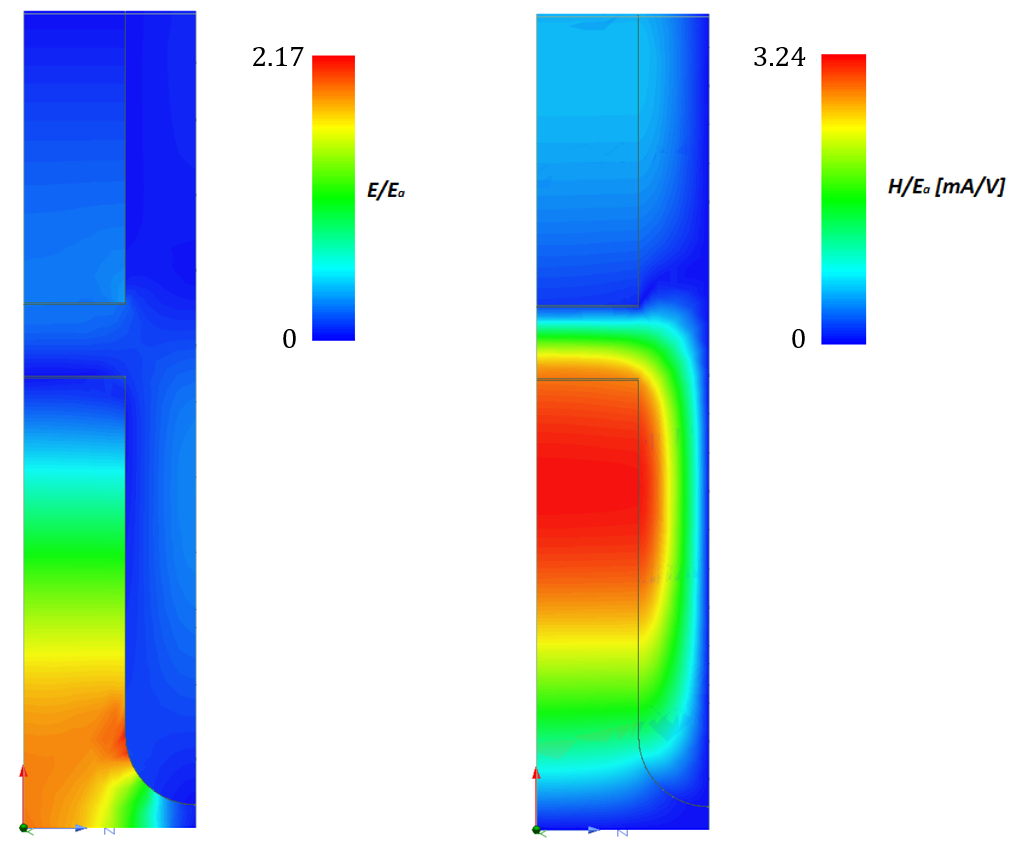}
\caption{Electric field distribution $E/E_a$ and magnetic field distribution $H/E_a$ for the accelerating mode TM$_{02}$-$\pi$ in a half regular cell using MgTiO$_3$, $\beta~=~0.6$ and $\xi=2$.}
\label{fig:field_distribution}
\end{center}
\end{figure}

The ratio of the peak electric field $E_p$ and the peak magnetic field $H_p$ to the average accelerating electric field $E_a$ usually limit the achievable accelerating gradient for conventional iris-loaded metallic structures, where

\begin{equation}
    E_a=\frac{1}{L_1}\int_{-L_1/2}^{L_1/2}E_z(0,0,z)\cos\left(\omega \frac{z}{\beta c}\right)dz,
\end{equation}
where $E_z$ is the longitudinal component of the electric field, $\omega=2\pi f$ is the angular frequency and $z$ is the longitudinal spatial coordinate. Field profiles for this structure are illustrated in Fig. \ref{fig:field_distribution}.

an optimization was done using four different ceramics, whose electromagnetic properties can be found in Table \ref{table:dielectric}, and particles velocity $\beta=\{0.4, 0.5, 0.6, 0.7, 0.8, 0.9, 1\}$. During these studies, it was observed that the geometry optimization depends mainly on the particle velocity $\beta$ and the relative electric permittivity $\varepsilon_r$ of the ceramic, while loss tangent $\tan \delta$ determines the final value of $Q_0$ as well as $Z_{eff}$. 

\begin{table}[htpb]
\centering
\caption{List of dielectrics studied in the optimization}
\begin{tabularx}{\linewidth}{XXXX}
\hline \hline
Material & Acronym & $\varepsilon_r$ & tan $\delta$ \\ \hline
CVD Diamond & Diamond & 5.7 & $3\times 10^{-6}$ \\
MgO & D9 & 9.64 & $6\times 10^{-6}$ \\
MgTiO$_3$ & D16 & 16.66 & $3.43\times 10^{-5}$\\
BaTiO$_x$ & D50 & 50.14 & $8\times 10^{-5}$\\ \hline \hline
\end{tabularx}
\label{table:dielectric}
\end{table}

\begin{figure}[h]
\begin{center}
\includegraphics[width=8cm]{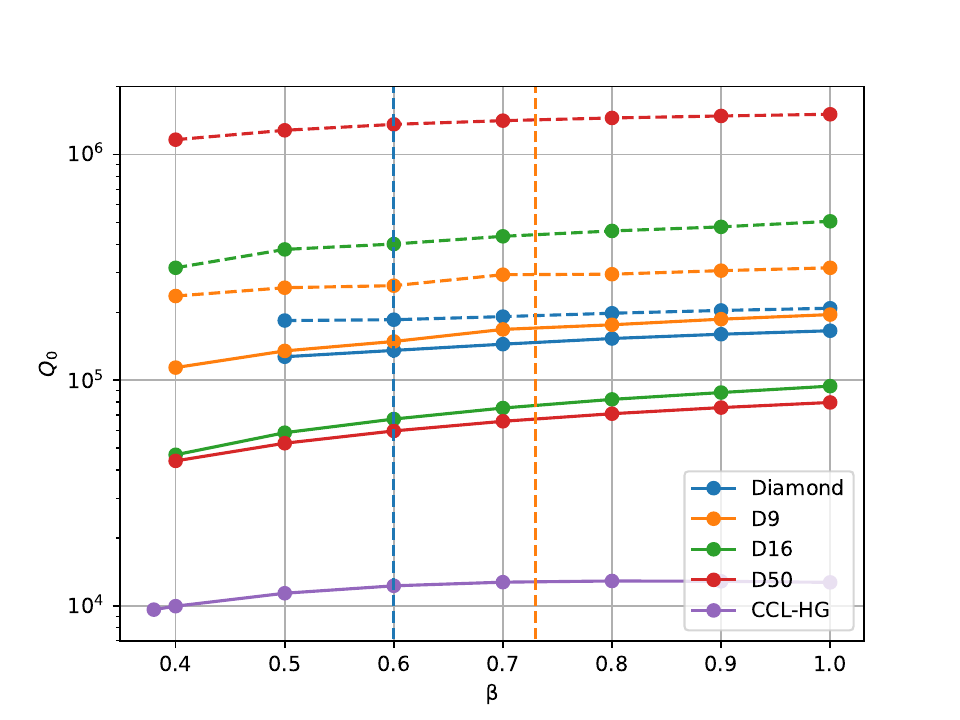}
\includegraphics[width=8cm]{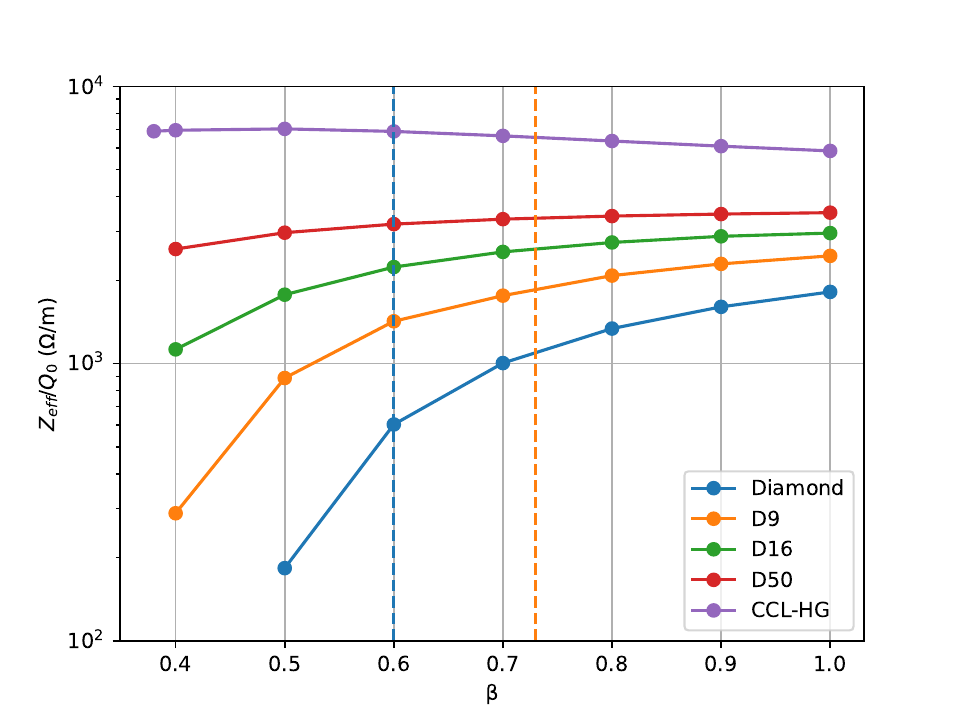}
\includegraphics[width=8cm]{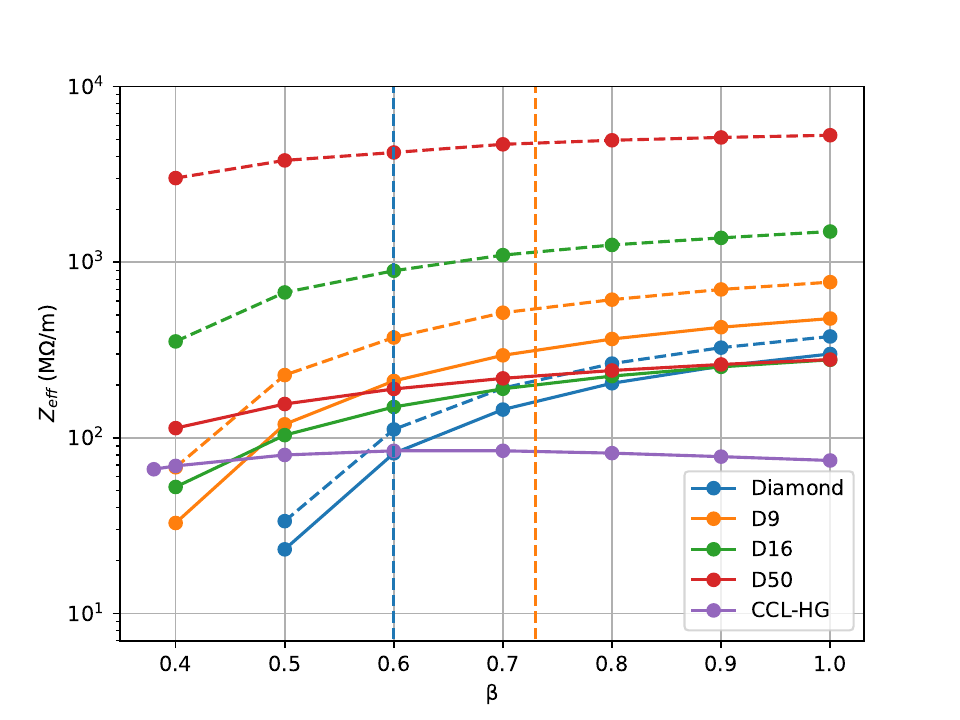}
\caption{Final results for $Q_0$, $Z_{eff}/Q_0$ and $Z_{eff}$ for the different values of particle velocity and different material in the ideal case (dashed lines), and taking into account dielectric losses (solid lines). Vertical dashed lines correspond to the maximum energy of protons (blue) and carbon ions (orange). The results are compared with a high-gradient cell coupled linac CCL-HG copper cavity (purple line) \cite{benedetti2017high}.}
\label{fig:beta_scan}
\end{center}
\end{figure}

Energy ranges for Hadrontherapy treatments vary between 70-230 MeV for protons and 100-430 MeV/u for carbon ions, which correspond to particle velocities between 0.37-0.60 and 0.43-0.73, respectively \cite{bencini2020design}. Final results for the figures of merit for these designs, taking into account dielectric losses, are compared with an extension for all particle velocities of a high-gradient standing wave copper cavity designed for protons with $\beta=0.38$ \cite{benedetti2017high}, as it can be seen in Fig. \ref{fig:beta_scan}.

Note the difference on the performance between the ideal and relativistic case. For the ideal case, the $Q_0$ is over two orders of magnitude higher compared to HG copper structures. As can be seen in Fig. \ref{fig:beta_scan}, the $Q_0$ increases with $\varepsilon_r$ and it is very sensitive to losses in the dielectric, though results are better than normal copper cavities. One caveat of this design is that a high amount of electrical energy is stored inside the dielectric, which is not going to be used to accelerate the beam. As a consequence, energy efficiency worsens, resulting in lower values of $Z_{eff}/Q_0$. Energy efficiency improves for larger $\varepsilon_r$ and, as expected, does not depend on the dielectric $\tan \delta$. The performance of the structure will be given by the shunt impedance, which is a compromise between $Q_0$ and $Z_{eff}/Q_0$. For the ideal case, the final result will be better for higher $\varepsilon_r$. However, due to the high sensitivity of $Q_0$ with dielectric losses, the characteristic value of $\tan \delta$ of the material is crucial on the real performance of the final structure. In addition, the performance increases also for higher particle velocities.
%and it is proved that high shunt impedance can be realised choosing the right dielectric.

\begin{comment}
It should be noticed that unloaded quality factor of DAA regular cell can achieved values over one order of magnitude higher than typical values of room-temperature copper-disk loaded cavities. Nevertheless, due to the their larger volume and the energy stored in the dielectric, the energy efficiency worsen, resulting in lower values for shunt impedance over quality factor, which does not depend on dielectric losses. Even though, it is possible to realise much higher shunt impedance using the right dielectric for the desired particle velocity. 
\end{comment}

\begin{figure}[h]
\begin{center}
\includegraphics[width=0.5\textwidth]{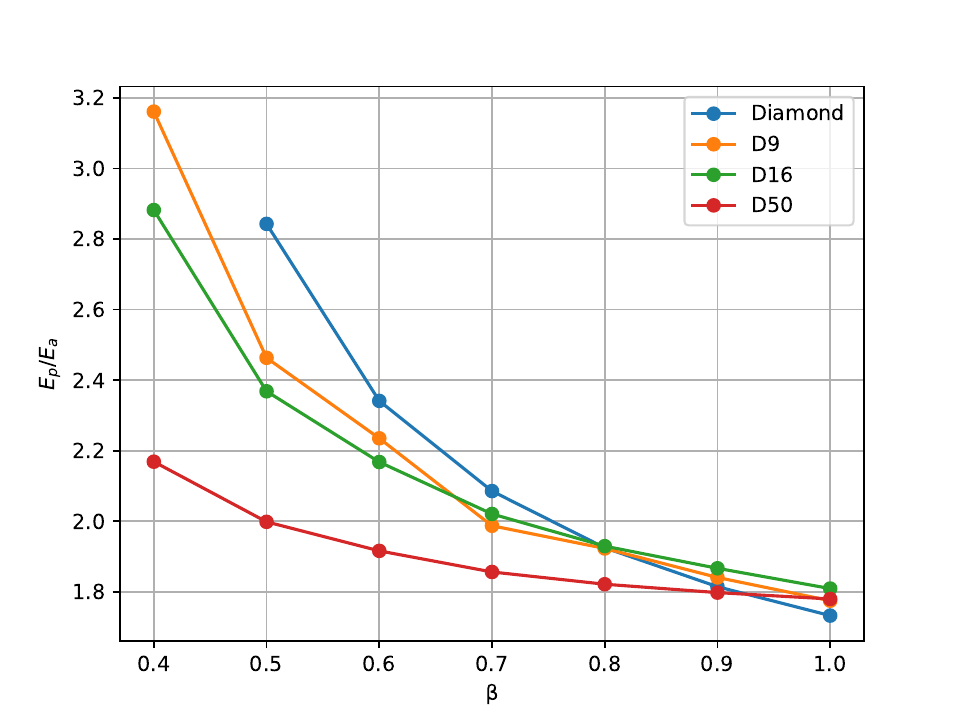}
\caption{Ratio of peak electric field and average accelerating field for different materials and particle velocities.}
\label{fig:E_ratio_betas}
\end{center}
\end{figure}

Moreover, as it is shown in Fig. \ref{fig:E_ratio_betas}, $E_p/E_a$ for the optimal geometry decreases with the material permittivity and particle velocity and it is always below 4 which is the value obtained for CCL-HG cavity \cite{benedetti2017high}. This ratio is one of the main limiting factors for HG cavities, since it is related with breakdowns production. Besides, it was observed that this ratio also decreases for thinner irises. Therefore, the performance of the DAA regular cell improves with particle velocity and higher electric permittivity being able to improve current values for room-temperature copper cavities.

\subsection{Dielectric loss tangent}
The advantage of working under the TM$_{02}$-$\pi$ mode, is that metallic losses are highly suppressed and, consequently, the DAA regular cell performance will be determined mainly by the quality of the dielectric in terms of its $\tan \delta$. 

Dielectric and metallic losses are given by \cite{satoh2016dielectric},
\begin{equation}\label{eqn:dielectric}
    P_d=\frac{1}{2}\omega \varepsilon_0\varepsilon_r\tan\delta  \int_V |\textbf{E}|^2d\tau,
\end{equation}
\begin{equation}\label{eqn:copper}
    P_c=\frac{1}{2}R_s\int |\hat{\textbf{n}}\times \textbf{H}|^2 dS,
\end{equation}
respectively, where $R_s=\sqrt{\omega\mu_0/(2\sigma)}$ is the surface resistance, $\varepsilon_0\varepsilon_r$ is the electric permittivity of dielectric, $\textbf{E}$ is the electric field, $\textbf{H}$ is the magnetic field, $\hat{\textbf{n}}$ is the unitary normal vector to the surface, $\mu_0$ is the magnetic permeability of vacuum and $\sigma=5.8\times10^7$~S/m is the copper electric conductivity. 

\begin{figure}[h]
\begin{center}
\includegraphics[width=8cm]{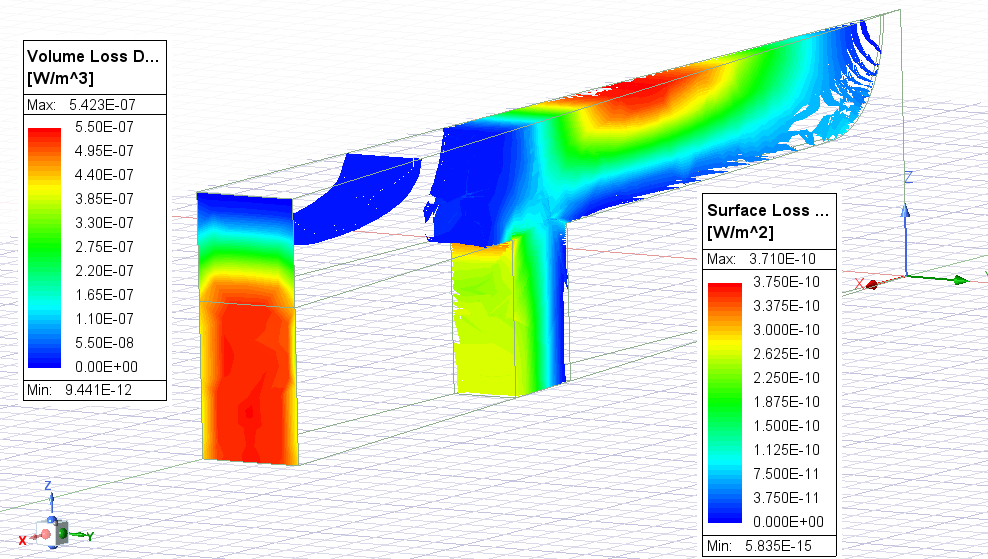}
\caption{Surface metallic and volumetric dielectric losses density for accelerating mode TM$_{02}$-$\pi$ in a regular cell using MgTiO$_3$, $\beta=0.6$, stored electromagnetic energy $W=1$ J and $\xi=2$.}
\label{fig:losses_distribution}
\end{center}
\end{figure}

\begin{figure}[h]
\begin{center}
\includegraphics[width=0.5\textwidth]{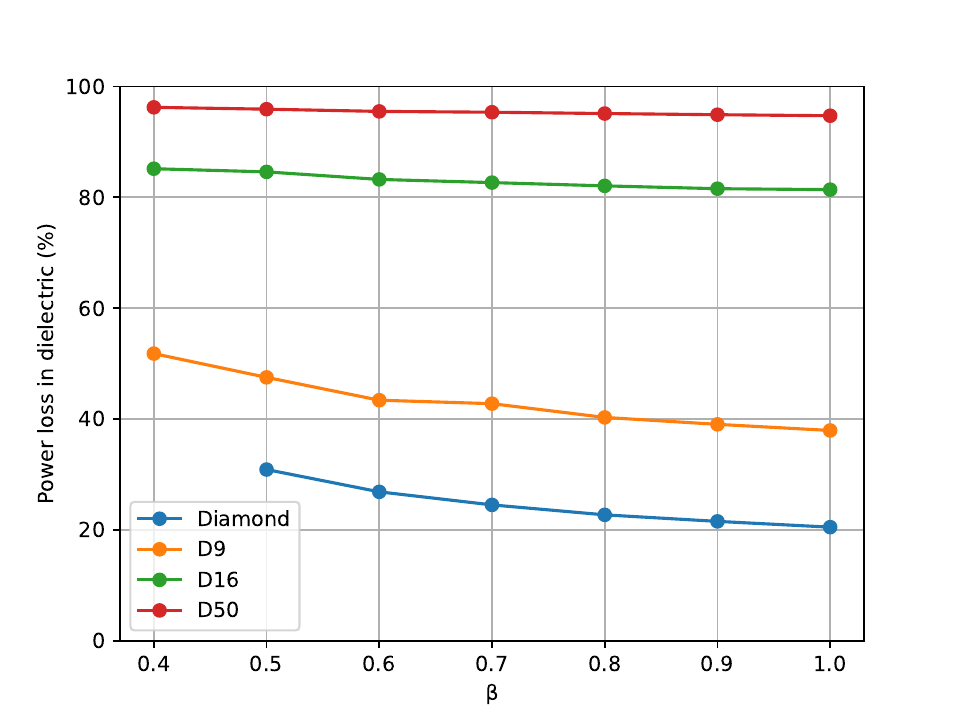}
\caption{Percentage of losses in the dielectric for different material and particle velocities.}
\label{fig:power_loss_dielectric}
\end{center}
\end{figure}

A graphical representation of both surface and volumetric loss densities are illustrated in Fig. \ref{fig:losses_distribution}. As both magnitudes have different units, a quantitatively comparison between them must be done by integrating the loss density over the whole geometry. Results are illustrated in Fig. \ref{fig:power_loss_dielectric}, showing a strong dependence mainly on the material loss tangent.

\begin{figure}[h]
\begin{center}
\includegraphics[width=0.5\textwidth]{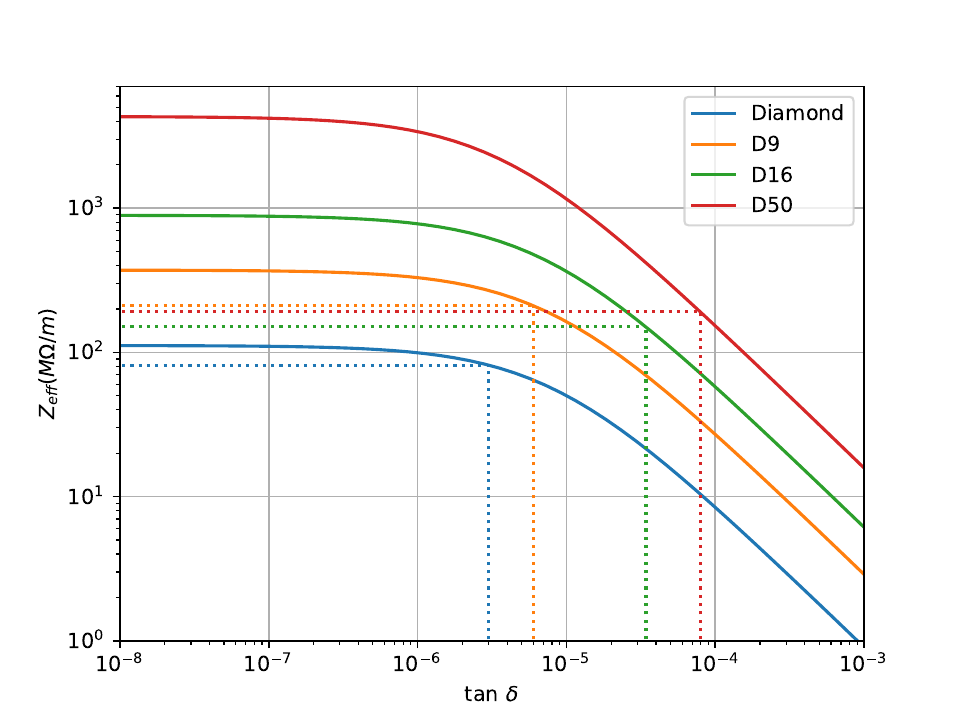}
\caption{Shunt impedance as a function of $\tan \delta$ for different materials and $\beta=0.6$. Experimental $\tan \delta$ values for each material are shown.}
\label{fig:losses_scan}
\end{center}
\end{figure}

As the loss tangent depends strongly on the manufacturing process for the ceramic fabrication, values of Table \ref{table:dielectric} are just references from previous experimental measurements. Therefore, it is of great importance to study the dependence of regular cell performance as a function of the $\tan \delta$ of the material, as illustrated in Fig. \ref{fig:losses_scan}, where an exponential decrease of the cavity performance can be seen for values of $\tan \delta >10^{-5}$.

\section{STRUCTURE PERFORMANCE}
Once the optimization of the geometry has been performed, the electromagnetic performance of the regular cell has to be considered as a component of a real accelerator system. In order to do so, the dispersion relation of the regular cell is studied as well as the field instabilities and singularities which can lead to multipactor or RF breakdown discharges. Besides, the consequences of using coating to suppress multipactor in the RF performance is also deliberated.

\subsection{Dispersion relation}
The overlapping between adjacent modes is a typical problem from the tunability and operational point of view for periodic RF accelerating structures, which is the case for the standing-wave accelerating structure studied in this work. In addition, good electromagnetic coupling between consecutive cells is also a key factor in order to determine the maximum number of cells per cavity.

\begin{figure}[h]
\begin{center}
\includegraphics[width=0.5\textwidth]{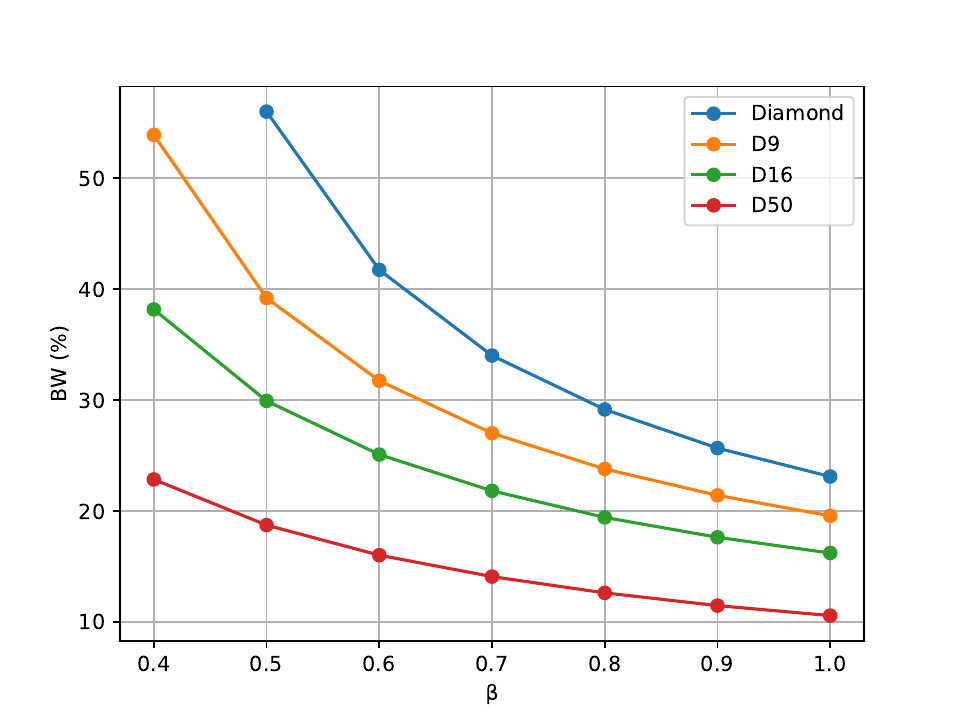}
\caption{Electrical coupling bandwidth for different materials and particle velocities with normalized iris thickness $\xi=1.5$. The bandwidth is defined as $BW=(f_\pi-f_0)/f_\pi\cdot 100\%$.}
\label{fig:coupling}
\end{center}
\end{figure}

Electric coupling between consecutive cells improves for lower electric permittivity and particle velocity, as it is illustrated in Fig. \ref{fig:coupling}. It can be observed that the TM$_{02}$ mode is strongly electrically coupled, so there is no need for coupling cells between regular cells.

% Numerical calculations showed that for thicker irises more modes are excited close to the resonant frequency of interest. This can lead to coupling of undesirable electromagnetic modes which are difficult to suppress due to its axial symmetry. In addition, electric coupling between consecutive cells improves for lower electric permittivity and particle velocity, as it is illustrated in Fig. \ref{fig:coupling}, implying that the coupling of the different modes will be more likely under these conditions. It can be observed that the TM$_{02}$ mode is strongly electrically coupled, so there is no need for coupling cells between regular cells. %However, as high coupling can lead to mode overlapping, RF ports must be carefully designed with low electric permittivity ceramics.

\begin{figure}[h]
\begin{center}
\includegraphics[width=0.5\textwidth]{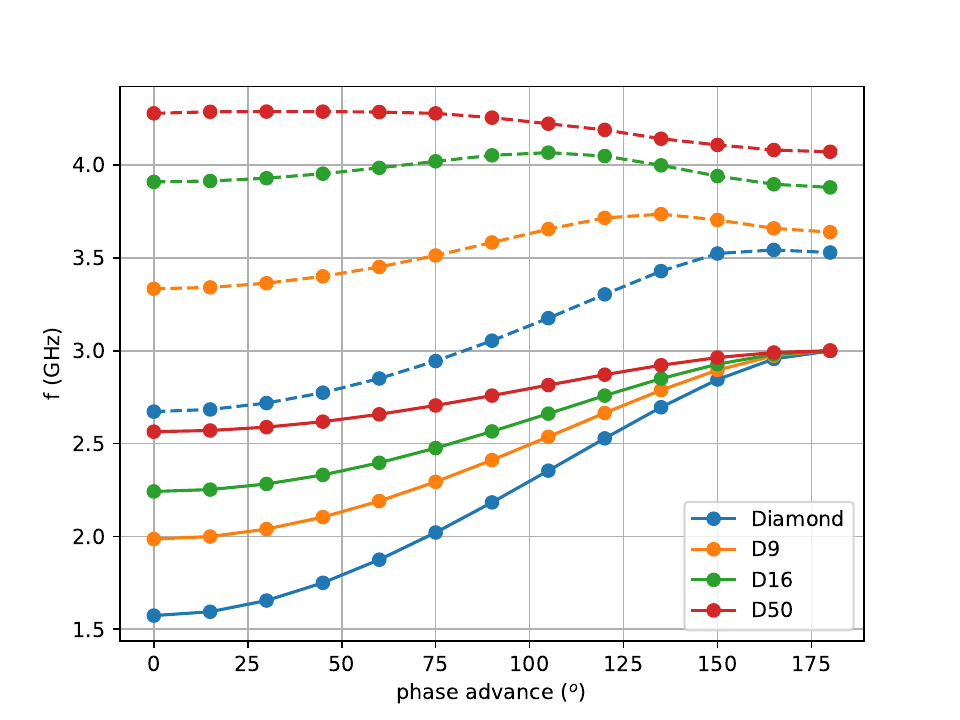}
\includegraphics[width=0.5\textwidth]{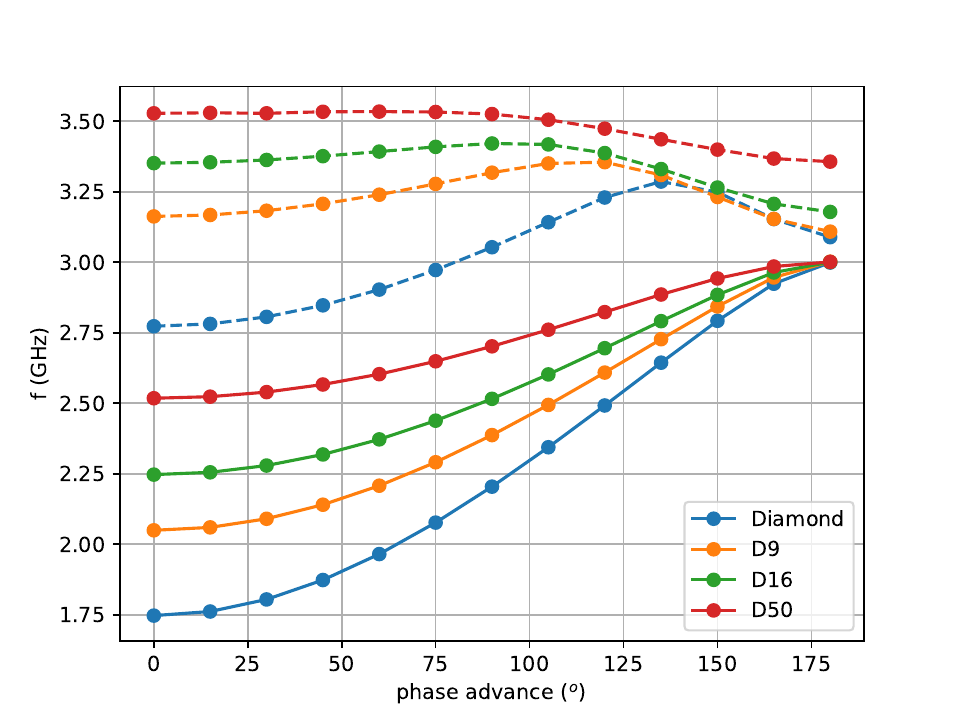}
\caption{Dispersion curve of the accelerating mode TM$_{02}$ (solid line) and next higher order mode (dashed line) for different dielectric materials and $\beta=0.6$ and normalized iris thickness $\xi=1$ (top figure) and $\xi=1.5$ (bottom figure).}
\label{fig:dispersion}
\end{center}
\end{figure}

Dispersion curves for the second order mode and the next higher order mode are depicted in Fig. \ref{fig:dispersion} for each material for two different normalized iris thickness. It can be seen that for low electric permittivity material, the higher order mode crosses the 3 GHz point and, therefore, overlapping cannot be avoided. In addition, as the iris becomes thicker, it can be seen that higher order modes with a phase advance of $\pi$ get closer to the resonant frequency and they can even cross this point for thicker irises. Consequently, electric permittivity and normalized iris thickness are bounded by the overlapping process.  
\begin{comment}
In addition, separation between different increases with the electrical permittivity of the material, which helps avoiding overlapping. 
\end{comment}

\subsection{Field instabilities}

The existence of sharp angles and triple junction points in the original design can lead to singularities in the surface electric field. As a consequence,  field instabilities, RF breakdowns or multipactor discharges can emerge.

\begin{figure}[h]
\begin{center}
\includegraphics[width=8cm]{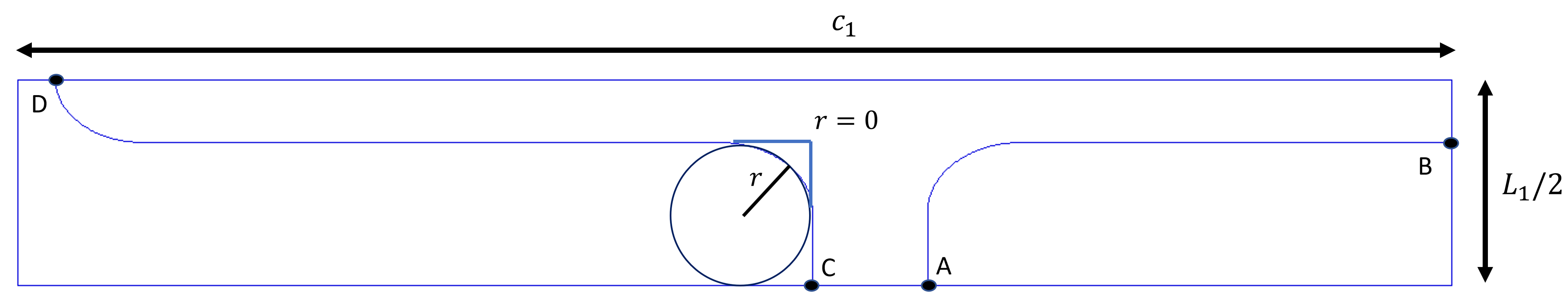}
\caption{Geometry modification.}
\label{fig:contorno}
\end{center}
\end{figure}

Regarding the triple junction point (point B in Fig. \ref{fig:contorno}), assuming zero conductivity in dielectric and flat metal wall, the electric field increases as $|\textbf{E}|\propto r^{n-1}$, where $r$ is the radial distance to the triple junction point and $n$ follows \cite{techaumnat2002effect}:
    \begin{equation}\label{eqtriple}
        \cot{n\alpha}+\varepsilon_r \cot{n(\pi-\alpha)}=0
    \end{equation}
    where $\alpha$ is the angle of vacuum between dielectric and metal and $\varepsilon_r$ is the relative electric permittivity of the dielectric.

From eq. \eqref{eqtriple} it can be concluded that if $\alpha<90$º then $n<1$, leading to infinitely large electric field in the junction, whereas if $\alpha>90$º then $n>1$, leading to null electric field. Only the case with $\alpha=90$º leads to $n=1$, implying a non-zero and non-singular value. However, we are just interested in avoiding singularities, which can be achieved by adjusting $\alpha\geq90$º. In addition, sharp metallic corners are another source of field singularities which must be avoided.

\begin{figure}[h]
\begin{center}
\includegraphics[width=0.5\textwidth]{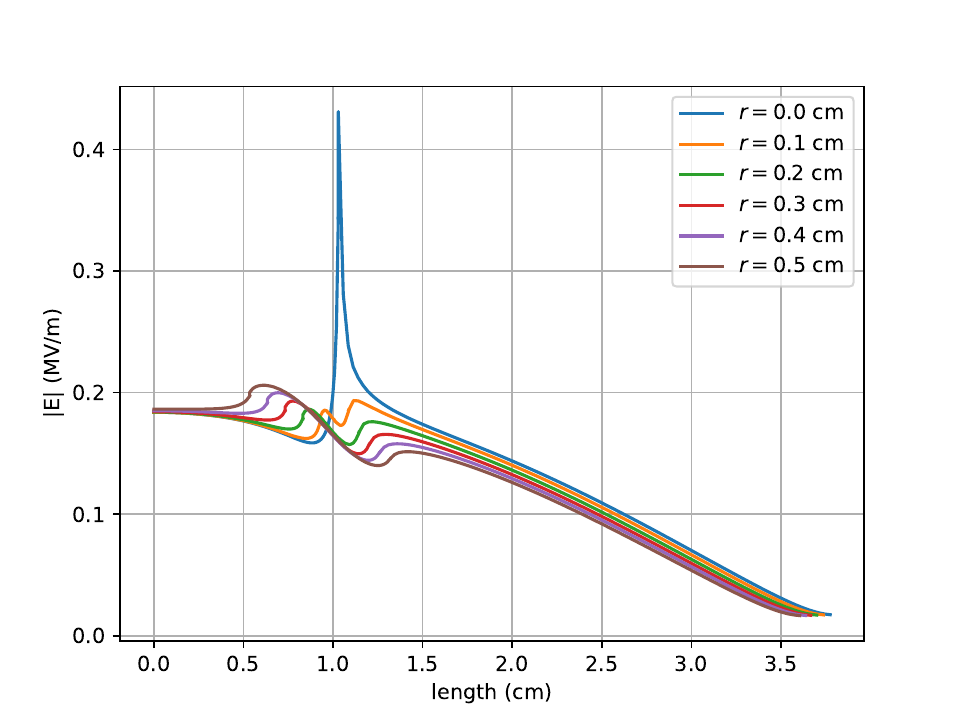}
\caption{Electric field magnitude along path \textit{AB} for different corner radius.}
\label{fig:AB}
\end{center}
\end{figure}

\begin{figure}[h]
\begin{center}
\includegraphics[width=0.5\textwidth]{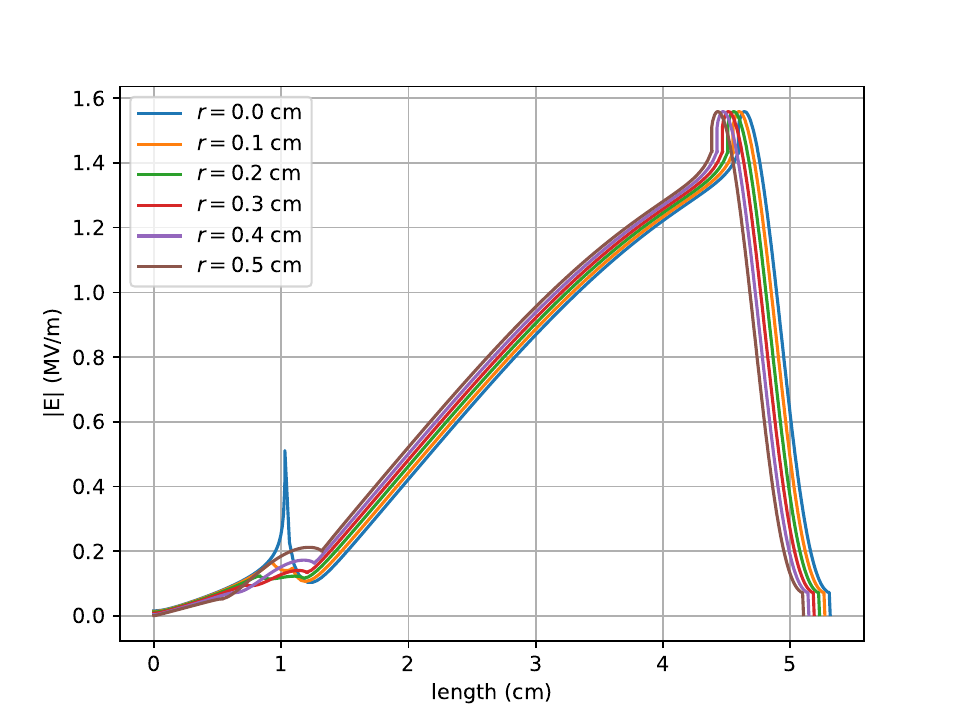}
\caption{Electric field magnitude along path \textit{CD} for different corner radius.}
\label{fig:CD}
\end{center}
\end{figure}

Regarding the dielectric corners in the junction between the dielectric ring and the iris, it was observed that sharp geometries also lead to field divergences. Thus, the geometry was changed as shown in Fig. \ref{fig:contorno} and the surface electric field for different round corners was studied for a fine mesh, as illustrated in Fig. \ref{fig:AB} and Fig. \ref{fig:CD}.

\subsection{Coating effects}

Amorphous Carbon (a-C) and Diamond Like Carbon (DLC) coatings were studied at Conseil Européen pour la Recherche Nucléaire (CERN) for Secondary Electron Yield (SEY) reduction in order to avoid multipactor discharges \cite{grudiev2022amorphous}. However, surface losses on the coating, which follow eq. \eqref{coating_losses}, will have an impact on the electromagnetic performance of the cavity

\begin{equation}\label{coating_losses}
    P_s=\sigma\int_V |\textbf{E}|^2 d\tau =\frac{1}{2R}\int|\hat{\textbf{n}}\times \textbf{E}|^2 dS,
\end{equation}
where $R$ is the sheet surface resistance of the coating.

\begin{figure}[h]
\begin{center}
\includegraphics[width=0.5\textwidth]{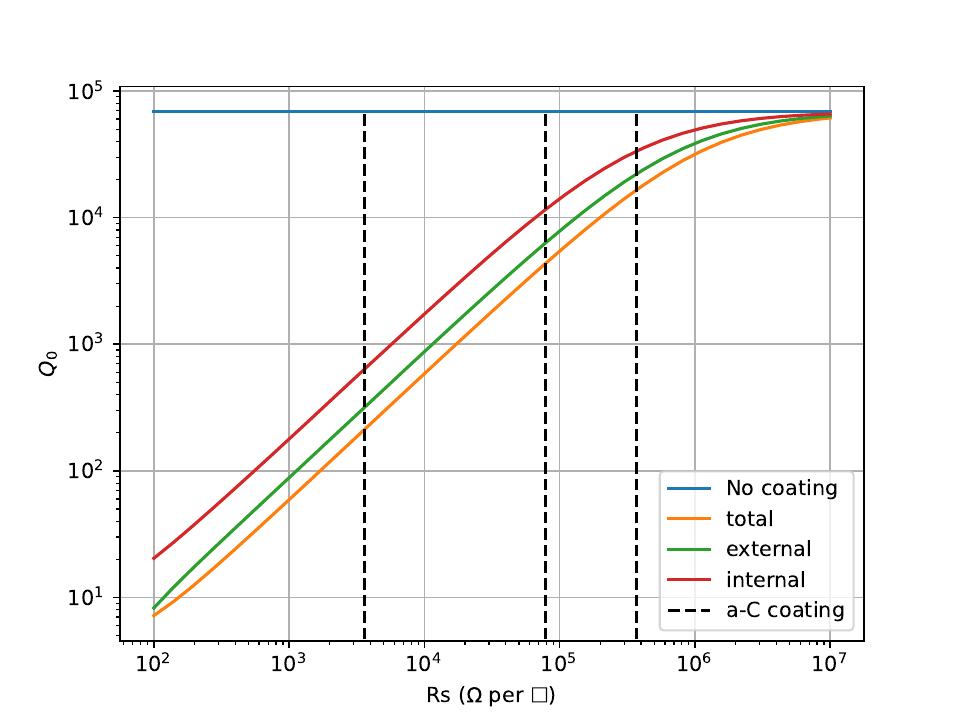}
\caption{Comparison of unloaded quality factor of a regular cell partially coated in the external or internal regions of the cavity, full covered and without coating as a function of the sheet resistance.}
\label{fig:coating}
\end{center}
\end{figure}

$Q_0$ as a function of the sheet resistance (in ohms per $\square$) is illustrated in Fig. \ref{fig:coating} for different cases: no coating, dielectric fully covered with coating, internal coating (which corresponds with coating in region \textit{CD}) and external coating (which corresponds with coating in region \textit{AB}). The surface resistance of DLC coating was above 1 M$\Omega$ per $\square$ and could not be measured, while a-C samples measurements are marked with black dashed lines.

As it can be observed in Fig. \ref{fig:coating}, $Q_0$ rapidly decreases for low resistance coatings and therefore thin films or materials with high resistivity are useful coatings to improve the electromagnetic performance.

\subsection{Thermal simulations}

In order to estimate the required cooling system and the mechanical stress and deformation induced by RF losses, thermal simulations were carried out using the ANSYS software \cite{ANSYS}. Volumetric and surface losses were used as input for steady thermal simulations with 3 cm of copper wall fixing the external temperature at 22ºC as boundary condition. Simulations were done for different geometries and ceramics, as illustrated in Table \ref{table:thermal}. Ultra high pure alumina was used for simulations instead of MgO because of its higher thermal conductivity in order to evaluate three different meaningful values.

\begin{table}[htpb]
\centering
\caption{List of dielectrics used for thermal simulations}
\begin{tabularx}{\linewidth}{XXXX}
\hline \hline
 Material & $\varepsilon_r$ & tan $\delta$ & $\kappa$  W·m$^{-1}$·K$^{-1}$ \\ \hline
CVD Diamond & 5.7 & $3\times 10^{-6}$ & 2000 \\
Al$_2$O$_3$ 99.99\% &  9.8 & $10^{-5}$ & 30 \\
MgTiO$_3$ &  16.66 & $3.43\times 10^{-5}$ & 3.8 \\\hline \hline
\end{tabularx}
\label{table:thermal}
\end{table}

\begin{figure}[h]
\begin{center}
\includegraphics[width=8cm]{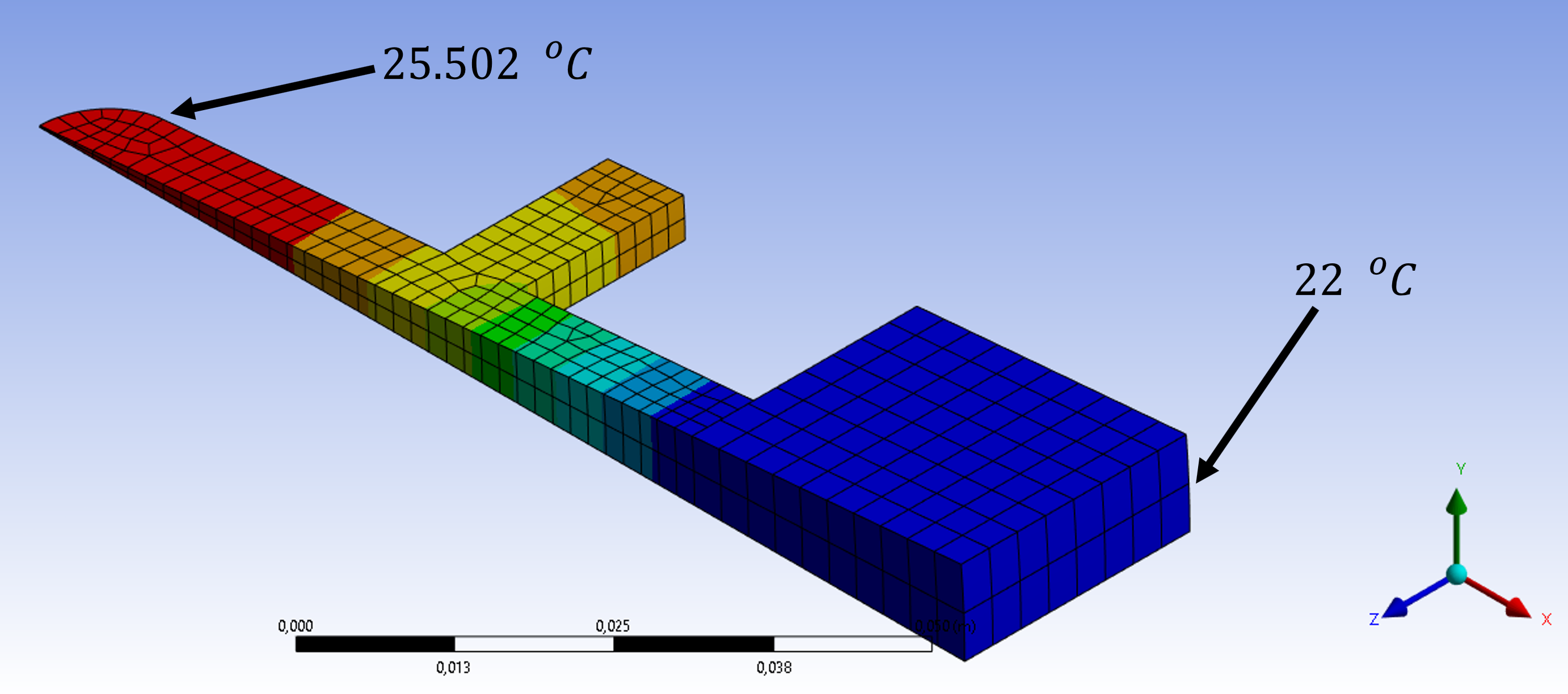}
\caption{Steady temperature distribution for Al$_2$O$_3$ and $\beta=1$.}
\label{fig:thermal}
\end{center}
\end{figure}

For numerical simulations, an accelerating gradient of 50~MV/m was used, with a duty cycle $D=0.075\times10^{-3}$,
\begin{equation}
    D=\frac{\tau}{T},
\end{equation}
where $\tau$ is the pulse width and $T$ is the total period of the signal. A graphical solution for Al$_2$O$_3$ for $\beta=1$ is shown in Fig. \ref{fig:thermal}.

\begin{figure}[h]
\begin{center}
\includegraphics[width=0.5\textwidth]{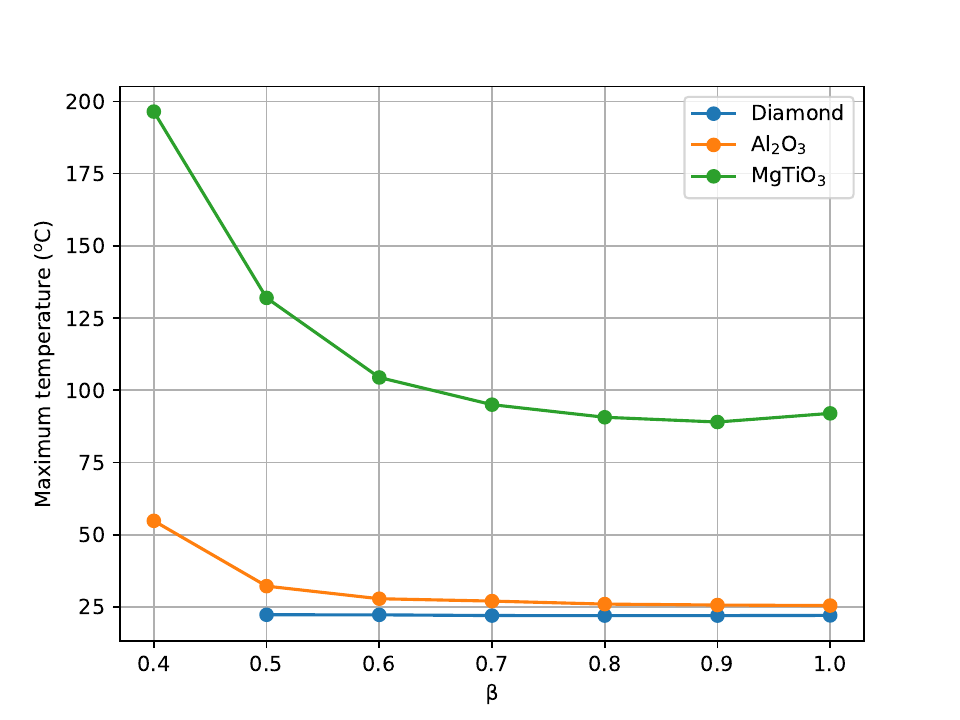}
\caption{Maximum temperature reached for different ceramics and particle velocities.}
\label{fig:temperature}
\end{center}
\end{figure}

As shown in Fig. \ref{fig:thermal}, the maximum temperature is reached close to the aperture of the ceramic, with a decreasing temperature gradient towards the copper metallic enclosure, which barely changes thanks to its high thermal conductivity ($\kappa=400$ W·m$^{-1}$·K$^{-1}$). The maximum temperature reached for different geometries and materials is illustrated in Fig. \ref{fig:temperature}. 

The temperature is higher for lower particle velocity while it seems to saturate at around $\beta=0.7$ following the behaviour of the $Z_{eff}$. In addition, very low thermal conductivity, as in MgTiO$_3$, leads to temperatures beyond acceptable limits regarding stress and deformation tolerances, even though we are still far from the fusion point.

\section{Conclusion and future work}
DAA structures for low $\beta$ particles have been studied for the first time, proving the potential to improve the performance of current room-temperature copper cavities. This study shows improvements in the optimization process and optimal results for an S-band DAA cavity as a solution for compact linear accelerators for Hadrontherapy treatments.

Working under the TM$_{02}$-$\pi$ mode, copper ohmic losses can be highly reduced by accumulating electrical energy inside the dielectric. From these studies we could conclude that cavity efficiency increases for higher particle velocity and electric permittivity. However, due to the high energy density inside the dielectric, the cavity performance will be limited by dielectric losses. Therefore, reaching low dielectric loss tangent is a fundamental key in the fabrication of ceramics in particular for DAA cavities.

Iris thickness plays also a fundamental role in the cell optimization by increasing the accelerating voltage and also by reducing the electric energy density inside the ceramic by decreasing dielectric losses. As a consequence, materials with higher loss tangent have thicker optimum irises than ideal geometries.

In addition, the ratio $E_p/E_a$ is lower than those reached by copper cavities, which potentially allows DAA cavities to reach higher gradients without producing RF breakdowns. This ratio decreases for high particle velocity, high electric permittivity and thin irises.

High electric coupling between consecutive cells has been observed for all kind of geometries. In addition, it was shown that coupling improves for lower particle velocity and lower electric permittivity. However low electric permittivity materials, such as CVD diamond, suffer from mode overlapping. In addition, thicker irises produce the excitation of more modes whose resonant frequencies are close to our operational frequency. As a result, the final design must find a compromise between an optimum electromagnetic design, which is achieved for thicker irises and low mode overlapping and low peak electric field, which improve for thinner irises.

Dielectric corners have been rounded in order to smooth the surface electric field. Moreover, stability studies of triple junction point were performed concluding that in order to avoid electric field singularities, the vacuum angle between dielectric and copper must be $\alpha\geq90$º and metallic sharp angles must be avoided.

Multipactor is one of the main limitations of DLA cavities due to high SEY of ceramics. Because of that, thin coating with low SEY is used for multipactor suppression. However, surface resistance of coating will have an effect on RF performance that must be studied in advance. Numerical simulations showed that low resistance coatings are unacceptable from an electromagnetic point of view, which implies that only high resistance materials or very thin coatings can be used in order to reduce multipactor.

Finally, thermal conductivity of the ceramic is found to be a crucial parameter also on the design to avoid overheating of the cavity leading to high deformation and stress. Thus, a lower bound value is set around 20-30 W·m$^{-1}$·K$^{-1}$ depending on particle velocity and duty cycle.

% if have a single appendix:
%\appendix[Proof of the Zonklar Equations]
% or
%\appendix  % for no appendix heading
% do not use \section anymore after \appendix, only \section*
% is possibly needed

% use appendices with more than one appendix
% then use \section to start each appendix
% you must declare a \section before using any
% \subsection or using \label (\appendices by itself
% starts a section numbered zero.)
%

\appendices

% use section* for acknowledgment
\section*{Acknowledgment}
Work supported by Ministerio de Universidades (Gobierno de España) under grant number FPU19/00585 and EST22/00739.

% Can use something like this to put references on a page
% by themselves when using endfloat and the captionsoff option.
\ifCLASSOPTIONcaptionsoff
  \newpage
\fi

% trigger a \newpage just before the given reference
% number - used to balance the columns on the last page
% adjust value as needed - may need to be readjusted if
% the document is modified later
%\IEEEtriggeratref{8}
% The "triggered" command can be changed if desired:
%\IEEEtriggercmd{\enlargethispage{-5in}}

% references section

% can use a bibliography generated by BibTeX as a .bbl file
% BibTeX documentation can be easily obtained at:
% http://mirror.ctan.org/biblio/bibtex/contrib/doc/
% The IEEEtran BibTeX style support page is at:
% http://www.michaelshell.org/tex/ieeetran/bibtex/
\bibliographystyle{IEEEtran}
% argument is your BibTeX string definitions and bibliography database(s)
\bibliography{references}
%
% <OR> manually copy in the resultant .bbl file
% set second argument of \begin to the number of references
% (used to reserve space for the reference number labels box)

% biography section
% 
% If you have an EPS/PDF photo (graphicx package needed) extra braces are
% needed around the contents of the optional argument to biography to prevent
% the LaTeX parser from getting confused when it sees the complicated
% \includegraphics command within an optional argument. (You could create
% your own custom macro containing the \includegraphics command to make things
% simpler here.)
%\begin{IEEEbiography}[{\includegraphics[width=1in,height=1.25in,clip,keepaspectratio]{mshell}}]{Michael Shell}
% or if you just want to reserve a space for a photo:

\begin{comment}
\begin{IEEEbiography}{Pablo Martinez-Reviriego}
Biography text here.
\end{IEEEbiography}

% if you will not have a photo at all:
\begin{IEEEbiographynophoto}{John Doe}
Biography text here.
\end{IEEEbiographynophoto}

\end{comment}

% insert where needed to balance the two columns on the last page with
% biographies
%\newpage

% You can push biographies down or up by placing
% a \vfill before or after them. The appropriate
% use of \vfill depends on what kind of text is
% on the last page and whether or not the columns
% are being equalized.

%\vfill

% Can be used to pull up biographies so that the bottom of the last one
% is flush with the other column.
%\enlargethispage{-5in}

% that's all folks
\end{document}